\documentclass[a4paper,11pt]{article}
\usepackage{graphicx}
\usepackage{amsfonts}
\usepackage{amsmath}
\usepackage{hyperref}
\title{Two closely related insecure noninteractive group key establishment schemes}
\author{Chris J. Mitchell\\Information Security Group, Royal Holloway, University of London\\
\url{www.chrismitchell.net}}
\date{7th March 2021}
\newcommand{\qed}{\nobreak \ifvmode \relax \else
      \ifdim\lastskip<1.5em \hskip-\lastskip
      \hskip1.5em plus0em minus0.5em \fi \nobreak
      \vrule height0.75em width0.5em depth0.25em\fi}
\begin{document}

\maketitle

\begin{abstract}
Serious weaknesses in two very closely related group authentication and group key establishment
schemes are described.  Simple attacks against the group key establishment part of the schemes are
described, which strongly suggest that the schemes should not be used.
\end{abstract}

\section{Introduction} \label{Intro}

In 2020 Cheng, Hsu and Harn proposed a combined (group) membership authentication and key
establishment scheme \cite{Cheng20} --- we refer to this scheme throughout as CHH\@.  The scheme is
claimed to be lightweight and hence suitable for wireless sensor networks (WSNs).  An extremely
similar scheme was then published by Hsu, Harn, Xia, Zhang and Zhao in early 2021 \cite{Hsu21} ---
we refer to this as the HHXZZ scheme. Rather disturbingly, although the Hsu et al.\ paper was only
submitted after the Cheng et al.\ paper had been accepted for publication and the two papers share
two authors, the Hsu et al.\ paper makes no reference to the earlier work.

There is a very extensive literature on group key establishment schemes, many of which at least
provide implicit authentication of the group members.  The interested reader is referred to Boyd,
Mathuria and Stebila \cite{Boyd20}.  It is far from clear whether, even it was secure (and it is
not, as we describe below), the two schemes offer any advantages over the state of the art, since
the only comparisons provided are with schemes using public key cryptography.

In this paper we describe a serious weakness which is shared by the two schemes. The remainder of
the paper is structured as follows.  In Section~\ref{sec-scheme} we briefly outline the operation
of the CHH scheme. An attack on the CHH scheme is described in Section~\ref{sec-attack}. The HHXZZ
scheme is then briefly introduced in Section~\ref{sec-scheme2}, and an attack is described which is
almost identical to the attack on CHH\@. Finally, concluding remarks are given in
Section~\ref{sec-conclusion}.

\section{The CHH scheme}  \label{sec-scheme}

The scheme involves a universally trusted \emph{Membership Registration Centre} (\emph{MRC}), which
provides information to each of $n$ participating entities $\{U_1,U_2,\ldots,U_n\}$.  This
information enables any subset of the entities to authenticate each other `as a group', and also to
establish a shared secret key which is not available to participating entities not in the subset.
The scheme uses arithmetic in GF($p$), the finite field of $p$ elements, for some prime $p>n$.  No
other requirements on $p$ are specified.

The scheme has five main stages, which we next briefly enumerate.  The first stage is used to set
up all the participants, and is only performed once.  The remaining four steps are performed
whenever a subset of entities wish to authenticate and establish a shared key. The reader is
directed to the Cheng et al.\ paper \cite{Cheng20} for the details --- the notation used below is
exactly as used in that paper.
\begin{description}
\item[0.  Token generation] This preliminary stage, performed once before active use of the
    scheme, involves the MRC generating and distributing a pair of `shares' ($s_i(y)$,
    $s_i(x)$) to each authorised participant $U_i$ ($1\leq i\leq n$), where $s_i(y)$ is a
    polynomial of degree $h-1$ over GF($p$) and $s_i(x)$ is a polynomial of degree $t-1$ over
    GF($p$), and where $h>2t-2$.
\item[1. Pairwise key generation] In this first operational stage, the members of a `group',
    i.e.\ a subset $\{U_{v_1},U_{v_2},\ldots,U_{v_m}\}\subseteq\{U_1,U_2,\ldots,U_n\}$, compute
    pairwise secret keys $k_{i,j}$ for each other using their shares.  In fact, this step could
    be performed just once as part of the initialisation process, since the pairwise keys will
    always be the same.
\item[2. Group authentication] This involves the members of the group mutually authenticating
    each other using the pairwise secret keys $k_{i,j}$.  After this step has completed each
    participant is confident that all members of the group agree on which entities are in the
    group.
\item[3. Group key establishment] This involves a further exchange amongst group members, as a
    result of which they agree on a shared secret key.  In this exchange, the value ($q_{v_i}$)
    sent by group member $U_{v_i}$ to all other group members is separately encrypted for each
    group member using the appropriate pairwise shared secret key (as established in step 1).
    The group key is then computed as the exclusive-or of the values
    $q_{v_1},q_{v_2},\ldots,q_{v_m}$ exchanged between group members.
\item[4. Group key authentication] This final stage, involving yet another exchange, is
    designed to give assurance that all members of the group agree on the shared secret key.
\end{description}

In the next part of this paper we describe an attack on the final two stages of the scheme, i.e.\
the group key establishment and group key authentication stages.

\section{An attack on CHH group key establishment}  \label{sec-attack}

\subsection{Some observations}  \label{sec-observations}

Before describing the attack, we make some minor observations on the operation of the scheme.
\begin{itemize}
\item There is no direct link between the group authentication stage and the group key
    establishment stage, except for the set of identities of the participants in the `group'.
\item The nature of the encryption function $E$ used in group key establishment is not
    specified. We assume here that it is instantiated as authenticated encryption (to avoid
    attacks that might be possible if encrypted values could be manipulated).
\item The scheme involves computing the bitwise-exclusive-or of values computed modulo $p$.  We
    assume here that prior to applying the exclusive-or operation the values are converted from
    integers to bit strings.
\end{itemize}

\subsection{Attack scenario, attack model and attack objective}

We suppose that a set of $m$ ($m\leq n$) participants $\{U_{v_1},U_{v_2},\ldots,U_{v_m}\}$ have
successfully completed the group authentication stage.

We further suppose that an (insider) adversary $U_{v_k}$ ($1\leq k\leq m$) controls the broadcast
channel with respect to `victim' participant $U_{v_j}$ ($1\leq j\leq m$, $j\not=k$), i.e.\ the
adversary can (a) prevent messages sent by other legitimate participants from reaching $U_{v_j}$,
and (b) send messages to $U_{v_j}$ on this channel that appear to have come from other legitimate
participants. Since the protocol makes no assumptions about the trustworthiness of the
communications channels, this assumption is legitimate.  Indeed, if the broadcast channel was
completely trustworthy, then much of the protocol would not be needed.

The objective of the adversary is to make the victim accept a key that is different to the key that
is accepted by all other members of the set $\{U_{v_1},U_{v_2},\ldots,U_{v_m}\}$. This would appear
to negate the purpose of the group key authentication stage, which is (presumably) all about
enabling all members of the `group' to verify that they share the same key.

\subsection{Subverting group key establishment}

The adversary $U_{v_k}$ first chooses a key $K^*$ which it wishes the victim $U_{v_j}$ to (wrongly)
accept as the shared group key. The adversary $U_{v_k}$ allows all messages sent by other
participants to reach their destinations correctly. However, the adversary sends two different
versions of its own message:
\begin{itemize}
\item it sends an encrypted version of the `correct' value $q_{v_k}$ to all participants
    $U_{v_s}$ ($1\leq s\leq m$) except for the victim $U_{v_j}$;
\item it sends an encrypted version of the value $q_{v_k}\oplus K\oplus K^*$ to the victim
    $U_{v_j}$, where $K$ is the `correct' shared group key.
\end{itemize}
Note that the adversary will need to wait until it has received all the values $q_{v_i}$ ($i\not=
k$) before it can send the value to the victim, since it must compute the group key $K$ before
sending the value.

As a result of the above steps, all participants except for the victim $U_{v_j}$ will share the
`correct' group key $K$.  However, the victim will believe that the group key is $K^*$.  We observe
in passing that:
\begin{itemize}
\item the adversary knows $K$ and $K^*$;
\item this part of the attack does \emph{not} require the adversary to manipulate the broadcast
    channel.
\end{itemize}

\subsection{Breaking group key authentication}

We conclude the attack by showing how the adversary can manipulate the authentication process so
that all participants believe the protocol has concluded successfully.  The authentication process
requires each participant to broadcast $H(K||L)$ where $H$ is a cryptographic hash function, $K$ is
the group secret key that has just been established, and $L$ is the sum of values broadcast (in
cleartext) at the beginning of the key establishment process.

To complete the attack the adversary needs to take control of the broadcast channel to and from the
victim $U_{v_j}$.  The victim will broadcast $H(K^*||L)$ --- the adversary suppresses this and
masquerades as the victim to broadcast $H(K||L)$.  All other participants will broadcast $H(K||L)$;
the adversary prevents these messages reaching the victim, and instead sends the victim `fake'
broadcasts of $H(K^*||L)$.

This completes the attack --- all participants except the victim will believe that $K$ is shared by
the group, and the victim will believe $K^*$ is shared by the group.

\section{The HHXZZ scheme and an attack}  \label{sec-scheme2}

\subsection{Operation}

The HHXZZ scheme is identical in operation to the CHH scheme except for step 4 (group key
establishment).  Even this step is very similar --- the only significant difference is in how the
group key is calculated from the set of values $\{q_{v_1},q_{v_2},\ldots,q_{v_m}\}$ exchanged
between group members (and how the values $q_{v_i}$ are calculated, although this makes no
difference to the attack so we ignore it here).

The HHXZZ scheme actually has two variants, one using addition and the other multiplication to
combine values. In \emph{Variant A} the group key is computed as
\[ K = \sum_{i=1}^m q_{v_i} \bmod p. \]
In \emph{Variant B} the group key is computed as
\[ K = \prod_{i=1}^m q_{v_i} \bmod p. \]

\subsection{Subverting group key establishment (again)}

The attack scenario, model and objective are precisely the same as for the CHH protocol.  We first
describe the attack for Variant A.

As previously, the adversary $U_{v_k}$ chooses a key $K^*$ which it wishes the victim $U_{v_j}$ to
(wrongly) accept as the shared group key. The adversary $U_{v_k}$ allows all messages sent by other
participants to reach their destinations correctly. However, the adversary sends two different
versions of its own message:
\begin{itemize}
\item it sends an encrypted version of the `correct' value $q_{v_k}$ to all participants
    $U_{v_s}$ ($1\leq s\leq m$) except for the victim $U_{v_j}$;
\item it sends an encrypted version of the value $q_{v_k}+K+K^* \bmod p$ to the victim
    $U_{v_j}$, where $K$ is the `correct' shared group key.
\end{itemize}

As a result of the above steps, all participants except for the victim $U_{v_j}$ will share the
`correct' group key $K$.  However, the victim will believe that the group key is $K^*$.  As
previously, this part of the attack does \emph{not} require the adversary to manipulate the
broadcast channel.

The attack for Variant B is exactly the same except that the adversary sends an encrypted version
of the value $q_{v_k}\times K^{-1}\times K^* \bmod p$ to the victim $U_{v_j}$, where $K$ is the
`correct' shared group key, and $K^{-1}$ is the multiplicative inverse of $K$ modulo $p$ (which is
easily computed using the Euclidean Algorithm).

Breaking the group key authentication step uses exactly the same procedure as for the CHH scheme.

\section{Concluding remarks}  \label{sec-conclusion}

We have demonstrated simple attacks which completely negate the security objectives of the
protocols. This means that the protocols should not be used.

Fundamentally, the fact that the authors have not provided rigorous proofs of security for the
schemes means that attacks such as those described here remain possible.  It would have been more
prudent to follow established wisdom and only publish schemes of this type if rigorous security
proofs had been established.  Similar remarks apply to the all-too-often misconceived attempts to
fix broken schemes, unless a proof of security can be devised for a revised scheme.  Achieving this
in an efficient way seems difficult for these schemes.

Finally, we observe that the two papers are extremely similar and build on precisely the same
(flawed) ideas.  The ethical issues raised by this are not discussed further here.

\providecommand{\bysame}{\leavevmode\hbox to3em{\hrulefill}\thinspace}
\providecommand{\MR}{\relax\ifhmode\unskip\space\fi MR }
\providecommand{\MRhref}[2]{%
  \href{http://www.ams.org/mathscinet-getitem?mr=#1}{#2}
} \providecommand{\href}[2]{#2}

\end{document}